\documentclass[epj]{webofc}
\usepackage[varg]{txfonts}   
\usepackage{graphics}

\woctitle{?????????}
\newcommand{\apjl}{\emph{Astrophys. J. Lett.} }

\newcommand{\app}{\emph{Astropart. Phys. }}

\newcommand{\prd}{\emph{Physics Rev. D } }

\begin{document}
\title{Future Extensive Air Shower arrays: from Gamma-Ray Astronomy to Cosmic Rays}
%
%

\author{Giuseppe Di Sciascio\inst{1}\fnsep\thanks{\email{giuseppe.disciascio@roma2.infn.it}} }

\institute{INFN, Sezione Roma Tor Vergata}

\abstract{Despite large progresses in building new detectors and in the analysis techniques, the key questions concerning the origin, acceleration and propagation of Galactic Cosmic Rays are still open. A number of new EAS arrays is in progress. The most ambitious and sensitive project between them is LHAASO, a new generation multi-component experiment to be installed at very high altitude in China (Daocheng, Sichuan province, 4400 m a.s.l.). The experiment will face the open problems through a combined study of photon- and charged particle-induced extensive air showers in the wide energy range 10$^{11}$ - 10$^{18}$ eV. In this paper the status of the experiment will be summarized, the science program presented and the outlook discussed in comparison with leading new projects.}
\maketitle
\section{Introduction}
\label{intro}

Cosmic Rays (hereafter CR) are the messengers of the \emph{'non-thermal Universe'}, generated in environments able to transfer energy and accelerate nuclei (and electrons) up to values not yet reached in our laboratories. 
Understanding their origin and propagation through the Interstellar Medium (ISM) is a fundamental problem which has a major impact on models of the structure and nature of the Universe.

The CR energy spectrum can be described by a power law with some features, the most important being the \emph{'knee'} around 3$\times$10$^{15}$ eV and the \emph{'ankle'} at about 10$^ {19}$ eV.
In the so-called standard model CRs below 10$^{17}$ eV are expected to be mainly Galactic, produced and accelerated in SuperNova Remnants (SNR), which can provide the necessary energy budget and naturally produce particles with a power law energy spectrum and spectral index close to the value inferred from the observations. The particles propagate diffusively in the ISM over very long time, under the effect of the Galactic Magnetic Fields (GMF) (for a review see, for example, \cite{blasi13} and references therein). 

The study of CRs is based on two complementary approaches:
\begin{itemize}
\item[(1)] Measurement of the energy spectrum, elemental composition and anisotropy of charged CRs.
\item[(2)] Search of their sources through the observation of neutral radiation (photons and neutrinos), which points back to the emitting sources not being affected by the magnetic fields.
\end{itemize}
The integrated study of charged CRs and of gamma rays and neutrinos, which should trace high energy hadronic interactions mainly nearby the still unidentified acceleration sites, is one of the most important (and exciting) fields in the so-called \emph{'multi-messenger astronomy'}.

Despite large progresses in building new detectors and in the analysis techniques, the key questions concerning origin, acceleration and propagation of CRs are still open.
In particular, the identification of sources in our Galaxy able to accelerate particles up to PeV energies, the so-called \emph{'PeVatrons'}, is still missing. In fact, even there is no doubt that galactic CR are accelerated in SNRs, the capability of SNRs to accelerate CRs up to the knee of the spectrum and above is still under debate.

In this paper we briefly summarize the motivation for new generation experiments addressed to study Galactic CR physics and $\gamma$-ray astronomy. We will focus on Extensive Air Shower (EAS)-arrays, not discussing new imaging Cerenkov projects like Cerenkov Telescope Array, described, as an example, in \cite{cta}.

\section{Open problems in Galactic Cosmic Ray Physics}
\label{sec-1}

In the knee energy region an accurate measurement of the CR primary spectrum can be carried out only by ground-based EAS arrays.
In fact, since the CR flux rapidly decreases with increasing energy and the size of detectors is constrained by the weight that can be placed on satellites/balloons, their collecting area is small and determine a maximum energy (of the order of 100 TeV/nucleon) related to a statistically significant detection. In addition, the limited volume of the detectors make difficult the containement of showers induced by high energy nuclei, thus limiting the energy resolution of instruments in direct measurements.

In the standard picture, mainly based on the results of the KASCADE esperiment, the knee is attributed to the steepening of the p and He spectra \cite{kascade}. 
However, a number of experiments reported evidence that the bending of the light component (p+He) is well below the PeV and the knee of the all-particle spectrum is due to heavier nuclei \cite{aglietta04,tibet,casamia,macro}. 
Interestingly, increasing the altitude of the experiment locations the measured average mass at the knee increases as well. 
As an example, recent results obtained by the ARGO-YBJ experiment (located at 4300 m asl) with different analyses reported evidence that the knee of the light component energy spectrum starts at $\sim$700 TeV, well below the knee of the all-particle spectrum that is confirmed by ARGO-YBJ at about 3$\cdot$10$^{15}$ eV \cite{discia-rev,discia14,demitri14,montini14}.

At higher energies, the KASCADE-Grande experiment reported evidence for a hardening slightly above 10$^{16}$ eV and a steepening at log10(E/eV) = 16.92$\pm$0.10 in the CR all particle spectrum.
A steepening at log10(E/eV) = 16.92$\pm$0.04 in the spectrum of the electron poor event sample (heavy primaries) and a hardening at log10(E/eV) = 17.08$\pm$0.08 in the electron rich (light primaries) one were observed. The slope of the heavy mass group spectrum above the knee-like feature is similar to the one of the light mass spectrum before the ankle-like feature \cite{kascade-grande}.

A large number of theoretical papers discussed the highest energies achievable in SNRs and the possibility that protons can be accelerated up to PeVs \cite{blasi13}.
The determination of the proton knee, as well as the measurement of the evolution of the heavy component across the knee, are the key components for understanding CR acceleration mechanisms and the propagation processes in the Galaxy, and to investigate the transition from Galactic to extra-galactic CRs.
As mentioned, experimental results in the knee range are still conflicting.
New high resolution measurements of the energy spectrum and elemental composition in the range from 10$^{12}$ to 10$^{18}$ eV are required to investigate in detail the role of different EAS observables and their dependence upon hadronic interaction models.

The measurement of the anisotropy in the arrival direction distribution of CRs is a complementary way to understand the origin and propagation of the radiation. It is also a tool to probe the structure of the magnetic fields through which CRs travel.

As CRs are mostly charged nuclei, their trajectories are deflected by the action of GMF they propagate through before reaching the Earth atmosphere, so that their detection carries directional information only up to distances as large as their gyro-radius. If CRs below 10$^{15}{\rm\,eV}$ are considered and the local GMF ($\sim3{\rm\,\mu G}$) is accounted for, gyro-radii are so short that isotropy is expected. At most, a weak di-polar distribution may exist, reflecting the contribution of the closest CR sources.

However, a number of experiments observed an energy-dependent \emph{``large scale''} anisotropy in the sidereal time frame with an amplitude of about 10$^{-4}$ - 10$^{-3}$, revealing the existence of two distinct broad regions: an excess distributed around 40$^{\circ}$ to 90$^{\circ}$ in Right Ascension (commonly referred to as ``tail-in'' excess) and a deficit (the ``loss cone'') around 150$^{\circ}$ to 240$^{\circ}$ in Right Ascension.
In recent years different experiments reported also the observation of a ``medium'' scale anisotropy in both hemispheres (for a review see \cite{disciascio13}).
Therefore, the observed pattern cannot be reproduced with a simple dipole, which implies the need at least two harmonics to fit data projected in sidereal time. 
In addition, results obtained by EASTOP and IceCube/IceTop showed that the anisotropy dramatically changes topology above 100 TeV and that the dipole component seems to disappear at PeV energies and beyond. 

From the theoretical viewpoint, the anisotropy of CRs is important as a direct trace of potential sources. Nonetheless, models failed to explain the whole set of CR anisotropy observations. 
So far, no theory of CRs in the Galaxy exists which is able to explain the origin of these different anisotropies leaving the standard model of CRs and that of the local GMF unchanged at the same time.
The anisotropy problem is the most serious challenge to the standard model of the origin of galactic CRs from diffusive shock acceleration \cite{hillas05}.

A measurement of the evolution of anisotropy with the energy across the knee and the determination of the chemical nature of excesses and deficits, are crucial to disentangle between different models of CR propagation in the Galaxy.
Moreover, the study of the anisotropy can clarify the origin of the knee. Indeed, if the knee is due to an increasing inefficiency in CR containment in the Galaxy a change in anisotropy is expected. If, on the contrary, the knee is related to the limit of the acceleration mechanism, we do not expect a change of anisotropy across the knee.

\section{Motivation for a Multi-TeV ($>$10 TeV) $\gamma$-ray detector}
\label{sec-2}

Actually TeV $\gamma$-rays have been observed from a number of SNRs, demonstrating that in SNRs some kind of acceleration occurs. However the question whether TeV gamma rays are produced by the decay of $\pi^0$ from protons or nuclei interactions (\emph{'hadronic'} mechanism), or by a population of relativistic electrons via Inverse Compton scattering or bremsstrahlung (\emph{'leptonic'} mechanism), still need a conclusive answer.

Recently AGILE and Fermi observed GeV photons from two young SNRs (W44 and IC443) showing the typical spectrum feature around 1 GeV (the so called \emph{'$\pi^0$ bump'}, due to the decay of $\pi^0$) related to hadronic interactions \cite{pizero-a,pizero-f}. 
This important measurement however does not demonstrate the capability of SNRs to accelerate CRs up to the knee of the spectrum and above. In fact, the identification of sources able to accelerate particles up to PeV energies, \emph{'PeVatrons'}, is still missing. 

In \emph{pp} interactions, 20\% of the proton energy goes into each pion flavor on average.
Therefore, each pionic gamma carries approximately 10\% of the energy of the proton primary. 
Thus the search for PeVatrons implies the observation of hundreds TeV photons.

The study of the multi-TeV ($>$10 TeV) emission is of particular importance in discriminating between different mechanisms.
A typical \emph{leptonic} (electron-produced) $\gamma$-ray spectrum would cut off earlier than typical \emph{hadronic} (CR-produced) spectra because of Klein-Nishina effects in the inverse-Compton cross section.
Therefore, a power law energy spectrum extending up to  hundreds TeV with no breaks can be reliably attributed to the hadronic mechanism.

To extend the energy range at PeVs a very large effective area is required.
The most sensitive experimental technique for the observation of multi-TeV $\gamma$-rays is the detection of EAS via large ground-based arrays.
The muon content of photon-induced showers is very low, therefore these events can be discriminated from the large background of CRs via a simultaneous detection of muons that originate in the muon-rich CR showers. 
Their large field of view (FoV) ($\Omega_{FoV}\sim$ O(sr)) and high duty cycle ($>$ 90\%) make these observational technique particularly suited to perform unbiased all-sky surveys (not simply of limited regions as the Galactic plane) and to monitor the sky for the brightest transient emission from Active Galactic Nuclei (AGN) and Gamma Ray Bursts (GRB), and search for unknown transient phenomena. 

Among 150 TeV sources, only less than 10 have been observed above 30 TeV, and the data are affected by large uncertainties. Even the most luminous TeV source, the Crab Nebula, is well studied only up to about 20 TeV.
In fact, for higher energies, the sensitivity of current instruments is not enough to determine clearly the spectral shape. 

It is known that the ensembles of the brightest sources in the GeV and TeV ranges do not coincide exactly, with some bright sources detected by Fermi relatively faint at TeV energy and vice-versa. It is therefore possible that the brightest sources in the 100 TeV ranges are unexpected. In this sense the large sky coverage of an EAS detector is well suited to discover the emission from unknown sources.
It has also been proposed, that the highest energy particle produced in astrophysical accelerators can escape rapidly the accelerator, and therefore that the highest energy emission is not from point-like or quasi point-like sources, but it could have a broader extension, mapping the interstellar gas distribution in the vicinity of the accelerators, and perhaps also the structure of the regular magnetic field in the region, if the diffusion is strongly anisotropic.
In these circumstances the identification of the PeVatrons is more difficult, but the wide FoV of the EAS-arrays could be decisive  to trace the $\gamma$-ray emission.

The space distribution of diffuse $\gamma$-rays can trace the location of the CR sources and the distribution of interstellar gas.
In fact, this emission can be produced by protons and nuclei via the decay of $\pi^0$ produced in hadronic interactions with interstellar gas.
In addition, the observation of a knee in the energy spectrum of diffuse galactic gamma-rays would provide a complementary way to investigate the elemental composition in the knee range.
The recent claim of a proton knee starting well below the PeV by the ARGO-YBJ experiment implies that a corrsponding knee in the diffuse energy spectrum is expected below hundred TeV and, in principle, could be detected by next generation $\gamma$-ray 
detectors.  Observing a location dependence of the knee energy would provide important clues on the nature of the knee, as do similar measurement for individual sources of CRs.

\begin{figure}
\centering
\begin{minipage}[ht]{.47\linewidth}
\vspace{-1cm}
  \centerline{\includegraphics[width=\textwidth]{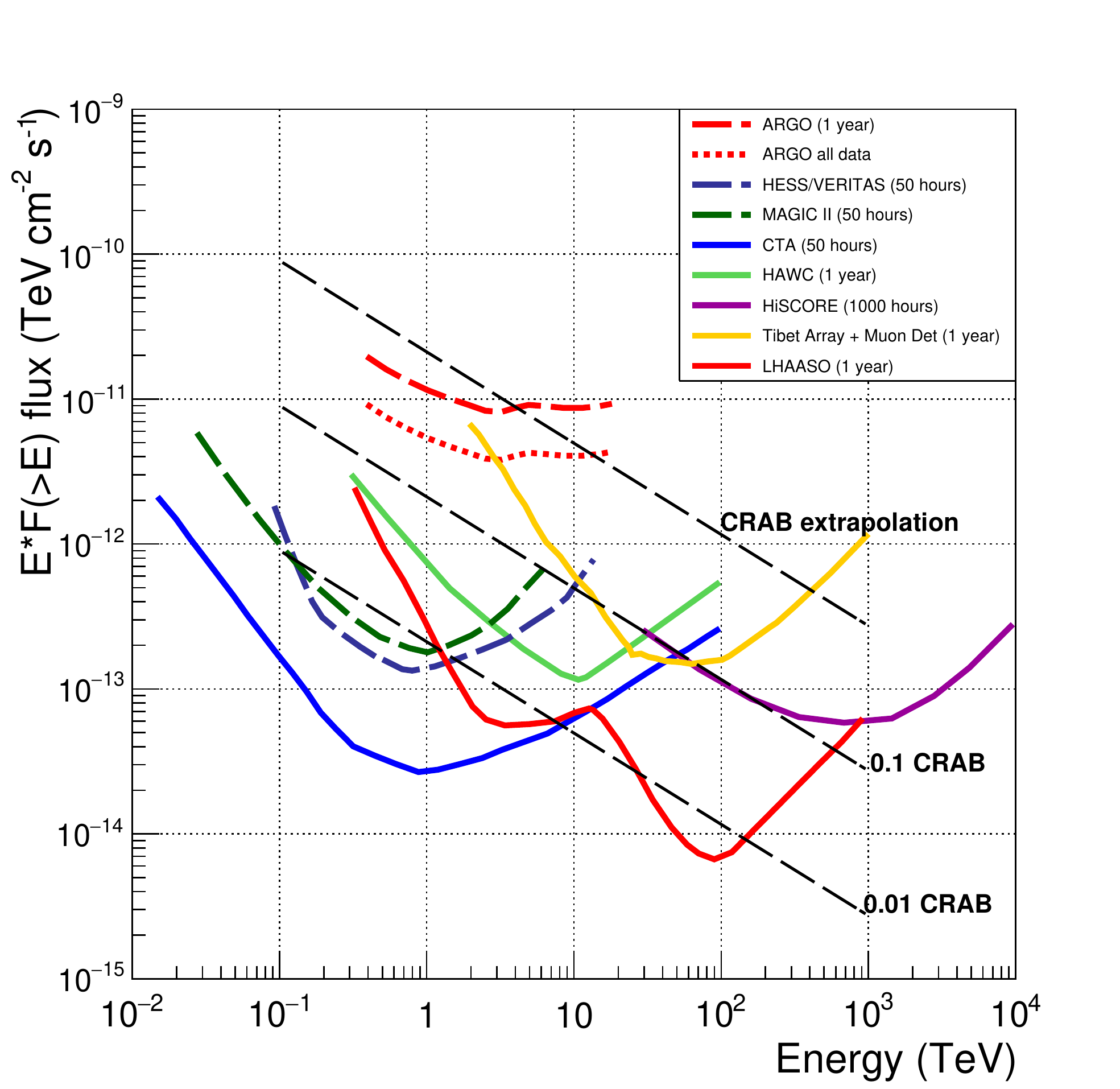}}
    \caption{Sensitivity of ground-based $\gamma$-ray detectors to a Crab-like point source.}
\label{fig:lhaaso_sens}
\end{minipage}\hfill
\begin{minipage}[ht]{.47\linewidth}
\vspace{-1cm}
  \centerline{\includegraphics[width=\textwidth]{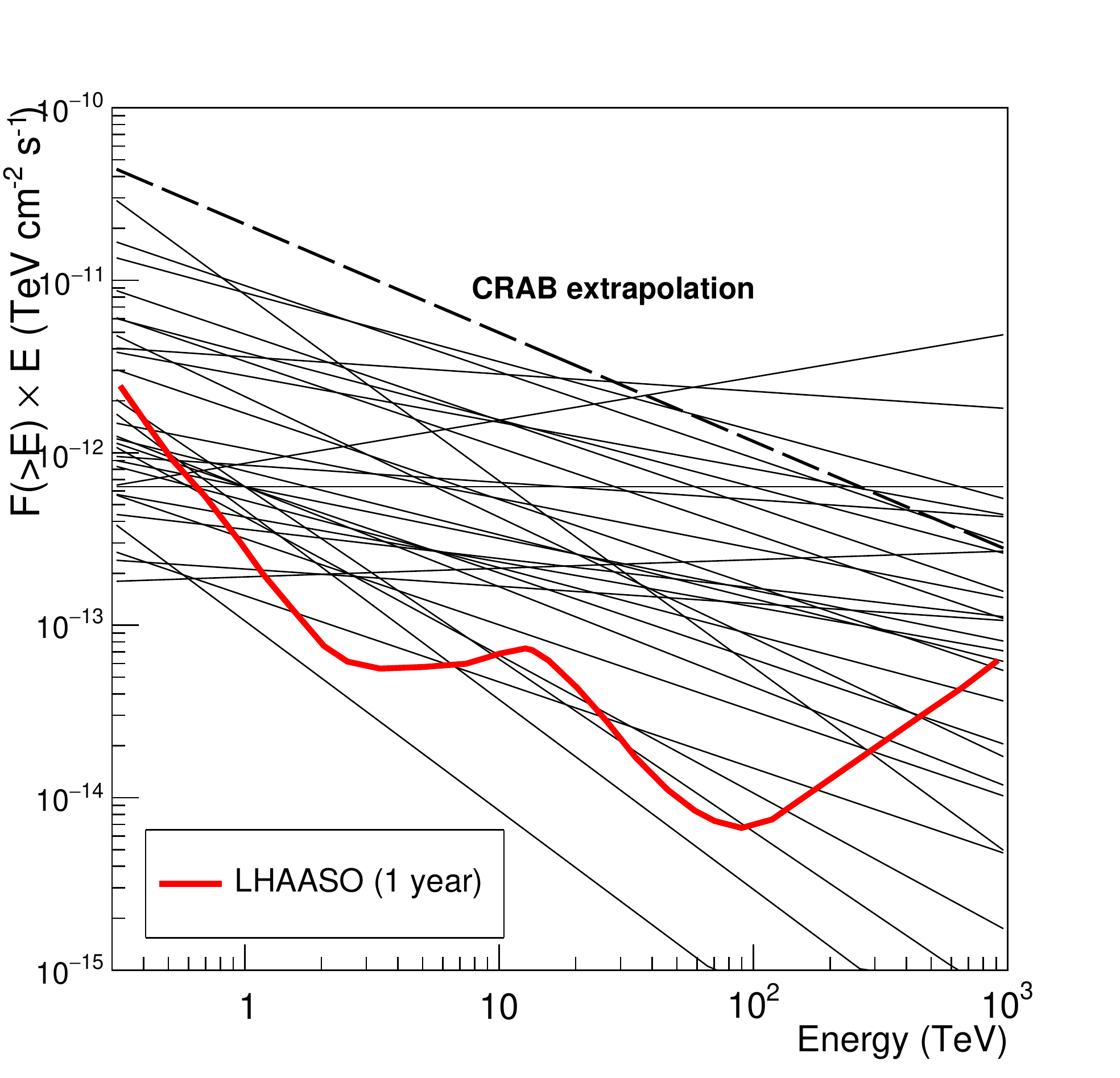} }
    \caption{Integral spectra of galactic $\gamma$-ray sources visible by LHAASO extrapolated to 1 PeV, compared to the LHAASO sensitivity (red curve) \cite{vernetto}.}
  \label{fig:gamma_sources}
        \end{minipage}\hfill
\end{figure}

\section{Future EAS-arrays}
\label{sec-3}

A number of new experiments or projects are under way in the coming years to study gamma-ray astronomy and CR physics. 
As mentioned, in this Section we focus on non-imaging Cherenkov and  particle detectors.

\subsection{Tibet AS$\gamma$}
The Tibet air shower array is located at Yangbajing (Tibet, P.R. China, 4300 m asl) and consists of (i) a high density array of over 500 scintillation counters with 7.5 m spacing, together with wider spaced outriggers, spread on a 370,000 m$^2$ instrumented area; (ii) a close packed array of 124 burst detectors distributed over a 500 m$^2$ area (YAC, Yangbajing Air shower Core detector array); (iii) 5 underground water Cerenkov muon detectors \cite{tibet-bd}.
The burst detectors are designed to detect the high-energy particles in the air shower core by converting them into electromagnetic cascade showers (burst) using a lead absorber 7 r.l. in thickness above a scintillation counter of the same area and 2 cm in thickness.
In the first phase, named "YAC-I", 16 YAC detectors each having the size 40 cm$\times$50 cm are distributed in a grid with an effective area of 10 m$^2$. YAC-I is used to check hadronic interaction models. The second phase of the experiment, called "YAC-II", consists of 124 YAC detectors. The inner 100 detectors of 80 cm$\times$50 cm each are deployed in a 10$\times$10 matrix 1.9 m apart and the outer 24 detectors of 100 cm$\times$50 cm each are distributed around them to reject events with the core outside the YAC-II array \cite{tibet-bd}. YAC-II is used to study the primary cosmic-ray composition in the range 5$\times$10$^{13}$ -- 10$^{16}$ eV.
The array has been operating in various configurations since 1990, and provides an angular resolution of $\sim$0.9$^{\circ}$ at 10 TeV. 
This experiment is mainly devoted to CR studies but the installation of muon detectors will allow the selection of high-energy ($>$ 10 TeV) photon-induced showers based upon the shower muon content. 
Data-taking with 5 of the planned 12 muon detectors and with YAC-II is started in 2014.

The expected sensitivity of Tibet AS$\gamma$ to point--like gamma ray sources \cite{tibet-sens} is shown in Fig. \ref{fig:lhaaso_sens} that plots, as a function of the threshold photon energy $E_{\rm min}$, the minimum integral flux $\Phi_\gamma (E_{\rm min})$ that is detectable with a 5~$\sigma$ statistical significance after one year (50 h) of operation for EAS-arrays (Cerenkov telescopes). The curve refers to the final layout with12 muon detectors.

\subsection{HiSCORE}

HiSCORE (Hundred*i Square-km Cosmic ORigin Explorer) will consist of an array of wide-angle ($\Omega\sim$ 0.6-0.85 sr) light-sensitive detector stations, distributed over an area of the order of 100 km$^2$. A HiSCORE detector station consists of four photomultiplier tubes, each equipped with a light-collecting Winston cone of 30$^{\circ}$ half-opening angle pointing to the zenith \cite{hiscore}. 
The primary goal of this non-imaging Cherenkov detector is gamma-ray astronomy in the 10 TeV to several PeV range. A prototype array of 9 wide-angle optical stations, spread on a 300$\times$300 m$^2$ area, has been deployed in the Tunka Valley near Lake Baikal since October 2013, and technical tests are underway. 
A 1 km$^2$ engineering array is planned for deployment in 2015, aiming at proof-of-principle measurements and first physics results.
The sensitivity of the HiSCORE standard configuration to point-sources is shown in Fig. \ref{fig:lhaaso_sens}. The curve has been calculated for 1000 h of exposure time, corresponding to 5 years of standard mode operation and roughly 1.4 years of tilted mode operation (see \cite{hiscore} for details).

\subsection{HAWC}

The HAWC (High Altitude Water Cherenkov Observatory) experiment, currently under construction at 4,100 m asl on the Sierra Negra volcano in Mexico, is the evolution of Milagro. The final array will consist of 300 water Cherenkov tanks of 7.3 m diameter and 4.5 m deep, covering an instrumented area of about 22,000 m$^2$ (the actual tank coverage is 12,550 m$^2$ with a coverage factor less than 60\%).
Deployment and commissioning are ongoing, with science operations already underway, and completion of the full array is expected in 2015.
Detector performance and first physics results are described in \cite{mostafa}.
The expected sensitivity to a Crab-like point-source is shown in Fig. \ref{fig:lhaaso_sens} \cite{hawc-sens}.

\section{The LHAASO project}
\label{sec-4}

LHAASO (Large High Altitude Air Shower Observatory) is a new generation instrument strategically built to act simultaneously as a wide aperture ($\sim$sr), continuosly-operated gamma ray telescope in the energy range between 10$^{11}$ and $10^{15}$~eV and as a high resolution CR detector in the broad energy range from 10$^{13}$ to 10$^{18}$~eV.

To achieve its  scientific  goals, the first phase of LHAASO will consist of the following major components \cite{lhaaso}:
\begin{itemize}
\item 1 km$^2$ array (LHAASO-KM2A), including 5635 scintillator detectors, with 15 m spacing, for electromagnetic particle detection.
\item An overlapping 1 km$^2$ array of 1221, 36 m$^2$ underground water Cherenkov tanks, with 30 m spacing, for muon detection (total sensitive area 40,000 m$^2$).
\item A close-packed, surface water Cherenkov detector facility with a total area of 90,000 m$^2$ (LHAASO-WCDA), four times that of HAWC.
\item 24 wide field-of-view air Cherenkov (and fluorescence) telescopes (LHAASO-WFCTA).
\item 452 close-packed burst detectors, located near the centre of the array, for detection of high energy secondary particles in the shower core region (LHAASO-SCDA).
\end{itemize}
LHAASO will be located at high altitude (4400 m asl, 600 g/cm$^2$, 29$^{\circ}$ 21' 31'' N, 100$^{\circ}$ 08'15'' E) in the Daochen site, Sichuan province, P.R. China.
The start of data taking is expected 2 -- 3 years after the start of installation planned at the beginning of 2016. The completion of installation in 6 years.

The sensitivity of LHAASO to point--like gamma ray sources is shown in Fig. \ref{fig:lhaaso_sens} where is compared to other experiments \cite{cui14}. The sensitivity curve has been calculated for a Crab Nebula like energy spectrum (power law with exponent $-2.63$) extending to PeVs without any cutoff. LHAASO is capable of observing sources with a brightness below $\sim$1~\% of the Crab flux in the energy ranges $\sim$1--10 TeV and $\sim$30--150 TeV. 
The LHAASO sensitivity curve shows a structure with two minima, reflecting the fact that the observation and identification of photon showers in different energy ranges is controlled by different components of the detector:  water Cherenkov detector (WCDA) in the range $\sim$0.3--10~TeV and KM2A array above 10 TeV. 

For comparison of sensitivities, it is important to note that the LHAASO sensitivity shown is the point-source survey sensitivity for $\sim$ 2 sr of the sky. 
While the sensitivities for EAS-arrays (ARGO-YBJ, HAWC, LHAASO, HiSCORE and Tibet AS$\gamma$) are also valid for surveys, the sensitivities for Cerenkov telescopes (CTA, HESS, MAGIC and VERITAS) are given for pointed observations of 50 h 'on source' in a small FoV of the order of $\pi$/100 sr. 
The choice of different conventions is inevitable because of the different operation modes of the two detection techniques.
Cherenkov telescopes  work only during clear moonless nights, with a total observation time of about 1000--1500 hours per year
(depending on the location), and have a FoV of a few degrees of radius. This implies that they can observe only one 
(or very few) sources at the same time, and only in the season of the year when the source culminates during night time.
Fifty hours is a typical time that a Cherenkov telescopes will dedicate to a selected source in one year. 

In contrast, the sky region observed by an EAS detector is completely determined by its geographical location.
The detector observes nearly continuously a large  fraction of the celestial sphere (spanning 360 degrees in right ascension and about 90 degrees in declination).
Sources located in this portion of the sky are in the FoV of the detector, either always, or for several hours per day,
depending on their celestial declination.
This situation is ideal to perform sky surveys, discover transients or explosive events (such as GRBs), and monitor variable or flaring sources such as AGNs.

One of the most interesting aspects of LHAASO is its  sensitivity to gamma rays above 30 TeV.
Inspecting Fig.~\ref{fig:lhaaso_sens} one can see that in this energy region LHAASO, thanks to the KM2A array,  will be the most sensitive gamma--ray telescope.
To give a more quantitative idea of the LHAASO performance, we can compare the detector sensitivities with the fluxes of the known TeV sources visible by LHAASO.
Fig. \ref{fig:gamma_sources} reports the spectra of galactic sources \cite{tevcat} extrapolated up to 1 PeV, using the same spectral slope measured in the TeV region, compared to the LHAASO one-year sensitivity \cite{vernetto}. 
The figure clearly shows that almost all fluxes are above the LHAASO sensitivity. Probably a consistent fraction of real sources spectra will not reach 1 PeV, but LHAASO will provide a unique information by studying in detail the spectral steepening or cutoff of galactic $\gamma$-ray sources.

As it is well known, an EAS detector is better suited than Cherenkov telescopes to survey large portion of the sky.
If we consider a survey of the Galactic Plane in the latitude band $-6^{\circ} < b < +6^{\circ}$, for a longitude interval of 200$^{\circ}$ (corresponding to the fraction of the Galaxy plane visible from the LHAASO site with zenith angle $\theta \lesssim 40^\circ$), in one year LHAASO can observe each source of this region for an average time of $\sim$1800 hours ($\sim$5 hours per day). 
A Cerenkov telescope, with its limited FoV, must scan the whole region with different pointings. The number of pointings determines the maximum observation time that can be dedicated to any source. 
Assuming a FoV of radius 5$^{\circ}$ with a decrease of sensitivity of about 50$\%$ at a distance of 3$^{\circ}$ from the center and a step for  different pointings  of $4^{\circ}$, the galactic survey requires approximately 150 pointings.
Assuming  a total observation  time of $\sim$1300 hours/year, a full year dedicated to the survey allows an  exposure of $\sim$9 hours for each galactic source.
Rescaling, as an example, the CTA sensitivity for an observation time of 9~hours, one finds that galactic survey performed by LHAASO in 1 year reaches a sensitivity better than CTA for E $\gtrsim$~ 2 TeV. The difference grows  with $E$, and is better a factor of $\sim$30 (170) for $E \gtrsim 30$ (80)~TeV.

The difference in sky survey capabilities between EAS-arrays and Cerenkov telescopes is more impressive in case of an {\it all sky survey}. Assuming a region of 7 sr to be scanned, the number of pointings is approximately $\sim$1400, allowing  only 0.9~hours/year of exposure per point in the sky.
Rescaling, as an example, the CTA sensitivity for an observation time of 0.9~hours, one finds that an all sky survey performed by LHAASO in 1 year is better than CTA by a factor of approximately 270 (1600) for $E \gtrsim 30$ (80)~TeV.

In conclusion, the two techniques (EAS detectors and Cherenkov telescopes) have both great scientific potential, and should be seen as complementary approaches in the exploration of different aspects, and different energy ranges, of the $\gamma$-ray emission.
On the contrary, CR studies are a peculiar characteristics of EAS-arrays

\section{Conclusions and outlook}
\label{sec-5}

The LHAASO experiment is expected to be the most sensitive detector to study the multi-TeV $\gamma$-ray sky opening the PeV range to the direct observations of the high energy CR sources.
The installation of an unprecedented muon detection area (40,000 m$^2$), together with arrays of wide field of view Cerenkov telescopes and burst detectors, will allow high resolution detection of nuclei-induced showers in the range 10$^{12}$ -- 10$^{18}$ eV.

As mentioned in the previous Section, all new wide field of view $\gamma$-ray experiments under way in the coming years will be located in the Northern hemisphere. 
To maximize the scientific return for Galactic sources, a future instrument should be located at sufficiently Southern latitude to continuously monitor the Galactic Center and the Inner Galaxy.
To lower the energy threshold down to hundred GeV range such an instrument would require very high altitude location (as an example, the 4800 m asl of the Alto Chorrillos region, hosting the Long Latin American Millimeter Array). The detection of very small low-energy showers with high efficiency and good angular resolution requires a detector exploiting a full coverage approach (with a coverage factor $>$90\%) with high segmentation of the readout. 
In fact, being precluded the $\gamma$/hadron discrimination due to the absence of muons in the sub-TeV showers, the angular resolution is the main parameter to improve the sensitivity.
The possible detection of PeVatrons in the Northern hemisphere by LHAASO (or HAWC) would require the increase of the collecting area with a large array.
In the near future such an instrument will be paired with the coming km$^3$ Mediterranean neutrino detector.

\end{document}